\documentclass[reprint, superscriptaddress, amsmath,amssymb, aps]{revtex4-2}
\usepackage{xcolor}
\usepackage{graphicx}% Include figure files
\usepackage{dcolumn}% Align table columns on decimal point
\usepackage{bm}% bold math
\usepackage{verbatim}
\usepackage[normalem]{ulem}

\usepackage[mathlines]{lineno}% Enable numbering of text and display math
\begin{document}

%\preprint{APS/123-QED}

\title{Spatial-temporal analysis of neural desynchronization in sleep-like states reveals critical dynamics}% Force line breaks with \\
%\thanks{A footnote to the article title}%

\author{Davor Curic}
 \email{dcuric@ucalgary.ca}
\affiliation{%
 Complexity Science Group, Department of Physics and Astronomy\\
 University of Calgary, Calgary, Alberta T2N 1N4, Canada 
}%
\author{Surjeet Singh}
\altaffiliation[Currently at: ]{Jackson Laboratory, Bar Harbor, ME 04609, USA}
\affiliation{
 Canadian Centre for Behavioral Neuroscience\\ University of Lethbridge, Lethbridge, Alberta T1K 3M4, Canada
}%

\author{Mojtaba Nazari}
\affiliation{
 Canadian Centre for Behavioral Neuroscience\\ University of Lethbridge, Lethbridge, Alberta T1K 3M4, Canada
}%

\author{Majid H. Mohajerani}
\affiliation{
 Canadian Centre for Behavioral Neuroscience\\ University of Lethbridge, Lethbridge, Alberta T1K 3M4, Canada
}%

\author{J\"orn Davidsen}%
\affiliation{%
 Complexity Science Group, Department of Physics and Astronomy\\
 University of Calgary, Calgary, Alberta T2N 1N4, Canada 
}%
\affiliation{Hotchkiss Brain Institute, University of Calgary, Calgary, Alberta T2N 4N1, Canada}

\date{\today}% It is always \today, today,
             %  but any date may be explicitly specified

\begin{abstract}
Sleep is characterized by non-rapid eye movement (nREM) sleep, originating from widespread neuronal synchrony, and REM sleep, with neuronal desynchronization akin to waking behavior. While these were thought to be global brain states, recent research suggests otherwise. Using time-frequency analysis of mesoscopic voltage-sensitive dye recordings of mice in a urethane-anesthetized model of sleep, we find transient neural desynchronization occurring heterogeneously across the cortex within a background of synchronized neural activity, in a manner reminiscent of a critical spreading process and indicative of an ‘edge of synchronization' phase transition.

\end{abstract}

%\keywords{Suggested keywords}%Use showkeys class option if keyword
                              %display desired
\maketitle

%\section{INTRODUCTION}

Mammalian sleep has largely been depicted as consisting of two basic states: non-rapid eye movement sleep (nREM), and rapid eye movement (REM) sleep~\cite{funk2016local}. A characteristic feature of nREM sleep is a high-amplitude electroencephalography signal with large low-frequency (delta band) power~\cite{buzsaki2006rhythms}, and is often referred to as slow-wave sleep. These large, slow signals are understood to occur due to a roughly one cycle per second synchronized oscillation of large populations of neurons~\cite{levenstein2019nrem}. Conversely, REM sleep is associated with a low-amplitude, higher-frequency (theta band) power~\cite{peever2017biology, lubenov2009hippocampal}. Unlike delta rhythms, theta rhythms are commonly seen in both sleeping and waking periods~\cite{buzsaki2006rhythms, lubenov2009hippocampal}, and in the cortex originate from neural desynchronization~\cite{mccarley2007neurobiology, funk2016local, peever2017biology, nunez2021theta}.

Both theta and delta rhythms have been observed in various regions of the brain, suggesting diverse functional purposes. E.g., cortical and hippocampal theta rhythms have been suggested to play a prominent role in memory formation~\cite{diba2007forward, schwindel2011hippocampal}. Despite this, the \emph{spatio-temporal} properties of these rhythms, and by extension that of neuronal de/synchronization, are not well understood, with many fundamental questions unanswered. Pertinent here, it was long believed that REM/nREM constituted global brain states --- that is, the brain as a whole occupies one of these two states, notable exceptions aside (e.g., unihemispheric sleep~\cite{mascetti2016unihemispheric, rattenborg2019local}). Yet, new research indicates that this global sleep hypothesis does not hold for terrestrial mammals such as mice~\cite{funk2016local, mohajerani2010mirrored, Nazari2022.03.01.481863}, and even humans~\cite{nir2011regional, bernardi2019regional, langille2019human}. Non-global wave-patterns of hippocampal theta activity has also been observed~\cite{lubenov2009hippocampal}. Indeed, there is even evidence of regionally localized slow-wave sleep during waking periods in both rodents and humans~\cite{mohajerani2010mirrored, bernardi2015neural}. All this indicates that both REM and nREM sleep constitute \emph{local} states, and their associated frequencies constitute localized, propagating rhythms~\cite{zhang2015traveling, muller2018cortical}. 

 %Using this method, we study the spatio-temporal evolution of a REM proxy signal and define desynchronization avalanches.

The challenge of studying spatial-temporal localization of neural rhythms has largely been a technological one. Measurement and bandwidth limitations enforce a trade-off between spatial and temporal resolution. E.g., probes can sample local field potentials (LFP) in the kiloHertz range, but offer limited spatial resolution. Only recently has the advent of faster computers, and new optical imaging techniques using fast florescent dyes~\cite{lecoq2019wide,grinvald2004vsdi}, made these studies feasible. Here, we employ wavelet transforms of voltage sensitive dye (VSD) data, obtained from wide-field imaging of the cortex of urethane-anesthetized mice, which serves as a model of sleep~\cite{clement2008cyclic}. Our analysis of spatio-temporal clusters of frequencies associated with desynchronization
suggests desynchronization spreads heterogeneously {in space and time } with no characteristic scale (i.e., scale-free), reminiscent of a critical mean-field directed percolation (MFDP) spreading process. This re-frames REM-like dynamics as a scale-free perturbation away from a synchronized state (nREM), providing strong evidence for neural dynamics operating between a totally synchronized and desynchronized phase --- i.e., an `edge-of-synchronization' phase transition --- opening a novel perspective on collective neural dynamics {in the form of \emph{desynchronization avalanches}}, and complementing on-going research into brain criticality and critical neuronal avalanches~\cite{beggs2003neuronal,di2018landau}.

Wide field imaging of VSD\cite{grinvald2004vsdi} data from a single cortical hemisphere was analyzed from 10 mice over 37 recordings (C57BL/6J mice RRID: IMSR$\_$JAX:000664, 6 six months old, 4 twelve months old), anesthetized with isoflurane (1.2$-$1.5$\%$) for induction, followed by urethane for data collection (1.0$-$1.2 mg/kg, i.p). For $in$ $vivo$ VSD imaging (RH1691 dye, Optical Imaging, New York, NY), a CCD camera imaged VSD fluorescence in an 8.6$\times$8.6 mm field of view, with 128$\times$128 pixels in total, and a 1 mm depth of field. See S1 for details on surgical procedure. Recordings were taken in 15 min epochs at 100 Hz. Hippocampal LFP from the pyramidal layer of dorsal CA1 was simultaneously recorded (20 kHz) using Teflon-coated stainless-steel wires (A-M Systems). Protocols \& procedures were approved by University of Lethbridge's Animal Welfare Committee and were in accordance with guidelines set forth by the Canadian Council for Animal Care. Hand-drawn masks were used to remove pixels not corresponding to the cortical surface. The remaining pixels were pre-processed (see S1~\cite{supp}) to extract VSD fluorescence changes, and filtered between 0.5-9 Hz.

\begin{figure*}[t!]
    \centering
    \includegraphics[scale = 0.47]{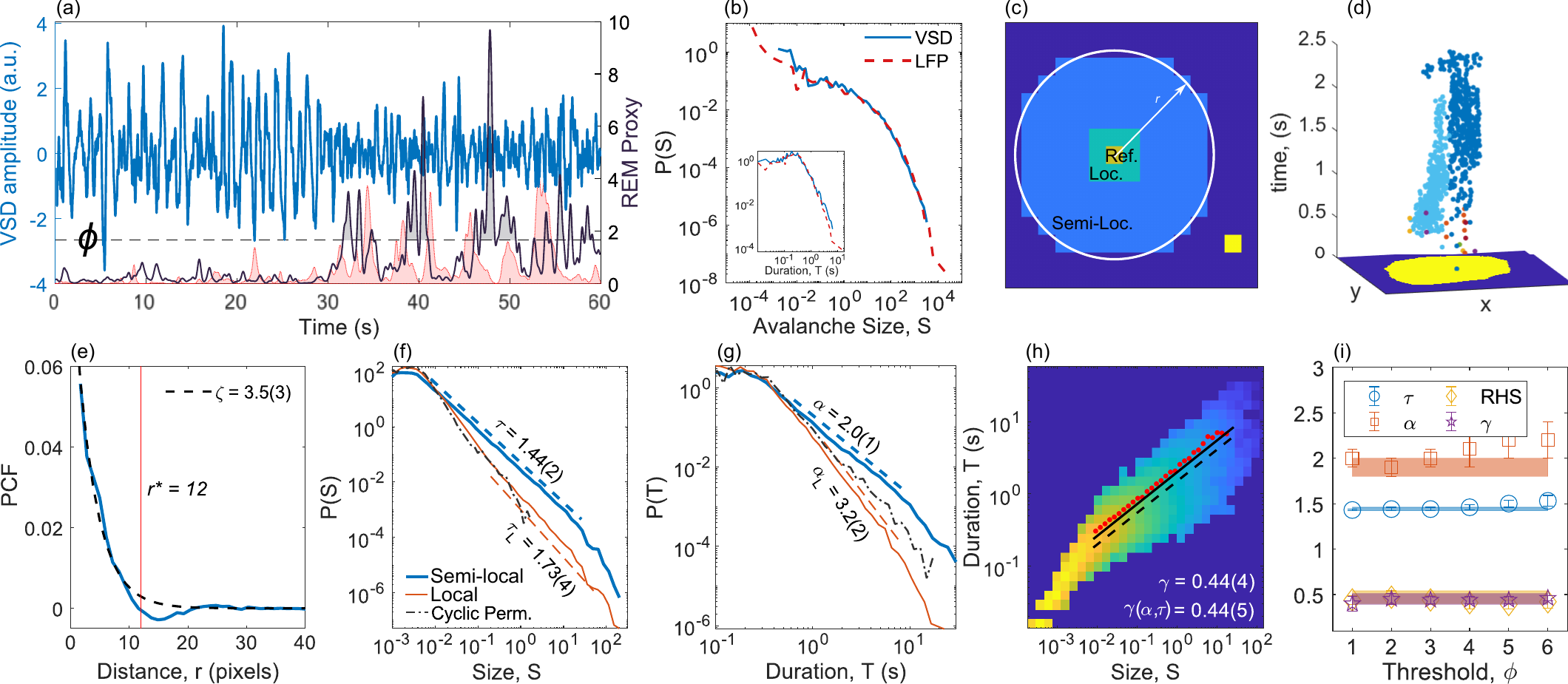}
    \caption{(a) Segment of the global VSD signal (blue) averaged over all pixels, and the corresponding REM proxy (black). Example of avalanches at a threshold $\phi$ (dashed) are grey-shaded. Red curve shows the hippocampal LFP REM proxy. (b) Avalanche sizes of the global VSD (blue) and hippocampal LFP (red, dashed) REM proxies ($\phi$ = 3 here and remaining panels). Inset shows avalanche durations.
    (c) Active sites within the semi-local neighbourhood of radius $r$ belong to the same avalanche as the reference site (Ref.). Loc. denotes the special case of local adjacency. In both cases the yellow pixel belongs to a separate avalanche, but may merge in the future.
    (d) Desynchronization avalanches (delineated by color) obtained using local adjacency. (e) Partial correlation function of a single recording (solid) with an exponential fit (dashed) with correlation length $\zeta$. Vertical line at $r^* = 12$ represents the first zero-crossing used for the semi-local analysis. (f) Avalanche size distributions obtained using semi-local (blue, thick) and local adjacency (red, thin). The slope and the analysis domain are indicated by colored dashed lines. The black dot-dashed curve corresponds to semi-local analysis of random cyclic permutation surrogates. (g) Same as (f) but for durations. (h) Density plot of avalanche durations against sizes for the semi-local analysis. Red dots denote the mean duration for that size. The solid line is $\gamma$ and the dashed line is obtained using Eq.~(\ref{eqn:scalingRelation}), and extends over the same domain as the blue dashed line in panel (f). (i) The semi-local exponents obtained across multiple thresholds, averaged over all animals. Error bars represent one standard deviation across all animals. Colored bars represent one standard deviation at $\phi = 3$.}
    \label{fig:dataAnalysis}
\end{figure*}

While anesthesia can differ from natural sleep, urethane was chosen as it has been reported to include nREM-REM transitions~\cite{clement2008cyclic, abadchi2020spatiotemporal}. Sleep states are traditionally classified using the ratio of theta band (5 - 9 Hz) power density to delta band (0.5 - 2 Hz) power density in LFP data~\cite{grosmark2012rem}. Yet, there is a strong correlation between cortical LFP and RH1691 signals such that one can use the latter as we do here~\cite{bermudez2018high}. High ratios indicate REM, whereas a low ratio corresponds to nREM. As such we refer to the ratio as the \emph{REM proxy}. As Fourier transforms cannot capture time-varying frequencies, wavelet transforms ($cwt$ MATLAB), which project onto temporally localized wave-packets (here Morse wavelets), are a more appropriate tool (Fig.~\ref{fig:dataAnalysis}(a))~\cite{hramov2015wavelets}. Lastly, results were robust against various definitions of theta and delta bands (see S2). Movies showing VSD and REM proxy dynamics on the cortical surface (amongst other metrics) are included in the supplementary material.

%\section{Analysis and Results}

%%\subsection{Global Avalanche Analysis}

First, we focus on desynchronization without spatial information by averaging the VSD signal over all pixels, then calculating the REM proxy signal. The signal at time $t$ is classified as desynchronized if the power density in the theta band is $\phi$ times greater than the delta band, and is classified synchronized otherwise. We place no condition on the duration of either state~\cite{koch2019automatic}. For $\phi=3$ we find on average across all mice $2.2\%$ (0.07$\%$, 6.7$\%$ are 5$^{th}$, 95$^{th}$percentiles) of the recording is classified as REM, which is consistent with other mice studies~\cite{mcshane2012assessing}. This can vary with age but no statistically significant differences were found between the 6 and 12 month old mice~\cite{soltani2019sleep}.

%Similar results were obtained for a range of thresholds and also if we use the REM proxy averaged over all individual pixels (not shown).
Desynchronization avalanches are defined as above-threshold REM proxy periods. We measure avalanche duration, $T$, and avalanche size, $S$, which is the area under the REM proxy relative to $\phi$~\cite{villegas2019time} (the grey-shaded area in Fig.~\ref{fig:dataAnalysis}(a)). As no significant differences were observed in avalanche statistics between mice (see S2), avalanches were joined across all animals for stronger statistics. Probability densities for $S$ and $T$, with $\phi = 3$, are shown in Figs.~\ref{fig:dataAnalysis}(b). No appreciable domain over which scale-free statistics are plausible could be established (i.e., Kolmogorov-Smirnov test with $p > 0.1$ for at least two orders of magnitude, see S3). These results were consistent for a range of thresholds ($\phi = 1, \ldots, 5$), also if we use the REM proxy averaged over all individual pixels (not shown). The hippocampal REM proxy is also shown in Fig.~\ref{fig:dataAnalysis}(a) (red, shaded), and largely follows the cortical proxy, suggesting the VSD analysis of REM-like sleep is meaningful. Moreover, avalanche statistics of the two are nearly identical, see Fig.~\ref{fig:dataAnalysis}(b).

Next, we focus on the \emph{spatio-temporal} properties of desynchronization. The REM proxy is computed for all pixels individually, and a global threshold $\phi$ delineates de/synchronization states, as before. A global threshold is chosen to avoid homogenization of any underlying spatial structure (see S6A), though avalanche statistics did not change when using a per-pixel threshold (S2). A spatio-temporal analysis requires a notion of spatial adjacency to be established. To account for longer ranged connections we define a `$semi$-$local$' adjacency - two above threshold pixels belong to the same avalanche if they are separated by at most one time-step, and within a given distance, $r^*$, of one another (Fig.~\ref{fig:dataAnalysis}(c)). The avalanche size is the number of constituent pixels across space and time, divided by the number of total pixels, $N$, (average  5.7(8)$\times 10^3$ pixels). A new feature is the coexistence and potential merging of multiple avalanches (Fig.~\ref{fig:dataAnalysis}(d)). If concurrent avalanches merge, the sizes are added and the duration is counted from the earliest avalanche~\cite{PhysRevX.11.021059}. We estimate $r^*$ using the partial correlation function, PCF (Fig.~\ref{fig:dataAnalysis}(e)). The partial correlation between two pixels controls for the activity of all others, measuring direct linear interactions (see S4). We chose the first zero crossing of the PCF at $r^* = 12$ as the radius for the semi-local analysis, which was found to be consistent across all animals, and avalanche statistics were consistent down to $r = 7$  (S4). We also consider `$local$' adjacency using the 8 neighboring pixels to establish the role of longer-ranged connections on avalanche statistics.

Figures~\ref{fig:dataAnalysis}(f,g) show avalanche size and duration distributions obtained using the semi-local and local analysis, and follow the power-law form $P(S) \propto S^{-\tau}$ and $P(T) \propto T^{-\alpha}$. Semi-local analysis of random cyclic permutation surrogates does not yield power laws (Fig.~\ref{fig:dataAnalysis}(f)), suggesting temporal coherence is an important factor. Exponents (and uncertainties representing 95$\%$ confidence intervals) were obtained using maximum likelihood estimation (MLE, $mle$, MATLAB), and scale-free statistics validated via alternate hypothesis testing (as in~\cite{clauset2009power}, see S3 for details). In the semi-local case, $\tau = 1.44(2)$ and $\alpha = 2.0(1)$, which are close to the critical MFDP expectations ($\tau$ = 1.5, $\alpha$ = 2)~\cite{munoz1999avalanche}. Conversely, the critical exponents of the local analysis ($\tau_L = 1.73(4)$ and $\alpha_L = 3.2(2)$) do not appear to match any well known universality class. The exponent $\gamma = 0.44(4)$, obtained by fitting to $\langle T \rangle(S) \propto S^{\gamma}$, is also close to the MFDP value of $0.5$ (Fig.~\ref{fig:dataAnalysis}(h)). At criticality the following scaling relation is expected to hold~\cite{PhysRevLett.122.208101}:
\begin{equation}
\frac{1}{\gamma} = \frac{\alpha - 1}{\tau - 1}.
\label{eqn:scalingRelation}
\end{equation}
Fig.~\ref{fig:dataAnalysis}(h) confirms this with $\gamma (\alpha,\tau) = 0.44(5)$.
%(uncertainty from standard error propagation). 
Avalanches also follow a symmetric universal temporal profile~\cite{friedman2012universal, hinrichsen2000non} regardless of duration (see S5). Importantly, estimated exponents are robust with respect to the threshold $\phi$ (Fig.~\ref{fig:dataAnalysis}(i)).
 
\begin{figure}
    \centering
    \includegraphics[scale = 0.48]{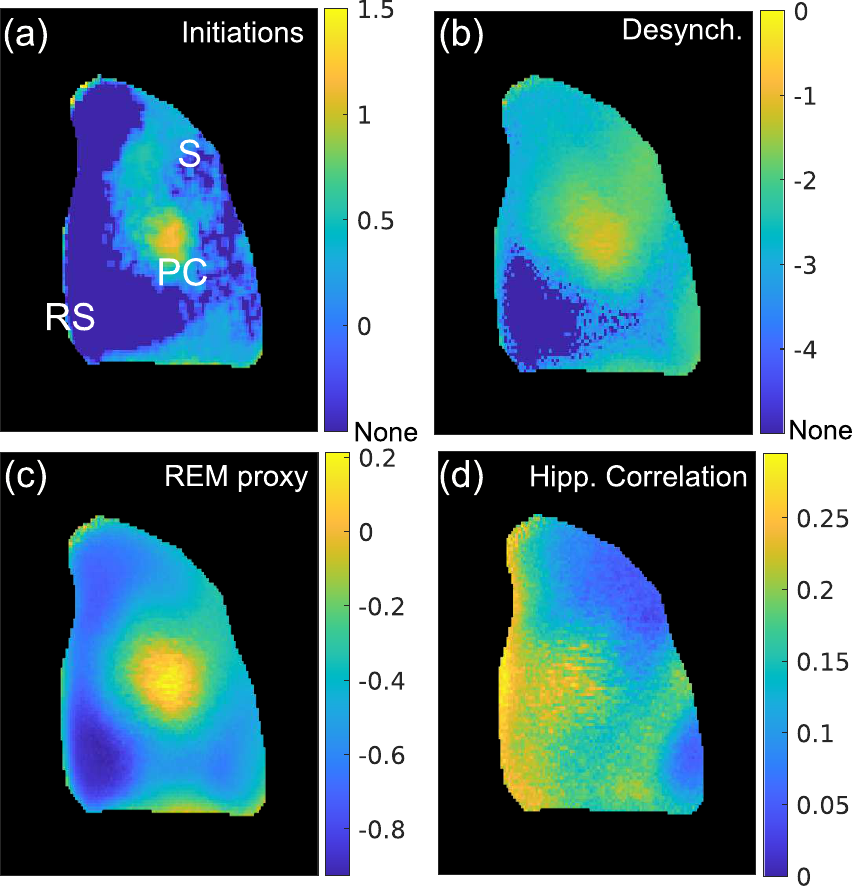}
    \caption{Density maps for a single mouse (colors are logscale, except (d)): (a) Average number of times a pixel initiated an avalanche ($\phi = 3$). The somatosensory (S), retrosplenial (RS), and parietal (PC) cortex are denoted. `None' indicates no events occurred for that pixel. (b) Number of times a pixel participated in an avalanche ($\phi = 3$), averaged over the total recording time. (c) Unthresholded REM proxy map averaged over the total recording time. (d) Cortical REM proxy correlation with the hippocampal REM proxy. 
    %Colors for all panels, except d, are logscale.
    } 
    \label{fig:spatialAnalysis}
\end{figure}

We now focus on the spatial features of desynchronization. Fig.~\ref{fig:spatialAnalysis}(a) shows (by representative example, all mice in S6) desynchronization avalanche initiations are detected heterogeneously across the cortical surface. We also find that the parietal cortex exhibits the highest rate of avalanching (Fig.~\ref{fig:spatialAnalysis}(b)), whereas the retrosplenial cortex (RSC) consistently had less avalanching. We hypothesize that the abundance of avalanching in the parietal cortex is related to its structural and functional hub nature, which has been established in rodents and humans under anesthesia \cite{lim2012vivo, coletta2020network, margulies2016situating}. While the RSC is also understood to be a structural hub~\cite{vann2009does}, Fig.~\ref{fig:spatialAnalysis}(c) shows this region has a low REM proxy relative to the rest of the cortex. Despite this, Fig.~\ref{fig:spatialAnalysis}(d) shows this region has higher correlations with the hippocampus, even within the theta band (see Fig. S5). The co-occurrence of high correlations and low REM proxy is possible since correlation analysis by definition standardizes signals and is, thus, amplitude independent, whereas avalanche analysis with a global $\phi$ is not (see, e.g., Fig. S8). When using a per-pixel threshold $\phi_i$ instead leaving correlations unaltered, many of the spatial features of the avalanches can no longer be detected by construction (see S6A).

\begin{figure}[t!]
    \centering
    \includegraphics[scale = 0.48]{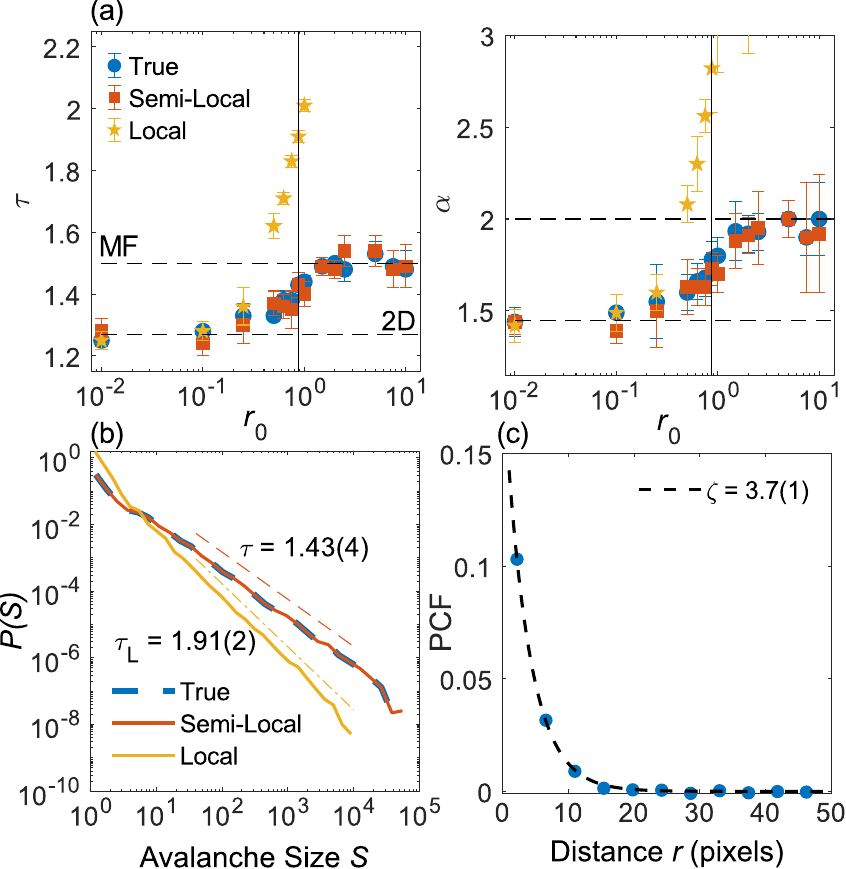}
    \caption{(a) Estimated size (left) and duration (right) exponents at $\sigma_c$ against the locality factor $r_0$ in Eq.~(\ref{eqn:propKernel}), with uncertainties from MLE. A transition is observed from mean-field (MF) to 2D directed percolation statistics (dashed lines). Vertical line denotes $r_0 = 0.85$ which best agrees with the experiments. (b) Avalanche size distributions. True avalanche distribution (dashed blue line), avalanches from the semi-local with $r^* = 12$ (red) and local analyses (yellow, subscript L) are shown. (c) Partial correlation function obtained using $r_0 = 0.85$ with the observed correlation length, $\zeta$, indicated.} 
    \label{fig:modelAnalysis}
\end{figure}

To establish whether a critical spreading process can capture the observed desynchronization dynamics given the connectivity structure within the cortex and its finite extent, we study a spreading process on a two-dimensional (2D) lattice --- with the same extent as the experimental field of view --- with varying spatial connectivity~\cite{PhysRevX.11.021059,schwartz2002percolation}. This is realized by initiating spreading from a randomly chosen single pixel $i$ and propagating the activity to sites $j$ with probability $p_{ij} = \sigma W_{ij}$, which in turn propagate the activity further. The parameter $\sigma$ tunes the dynamics to criticality and
\begin{equation}
W_{ij} = \exp(-r_{ij}/r_0).
\label{eqn:propKernel}
\end{equation}
Here, $r_{ij}$ is the (pixel) distance, and  $r_0 > 0$ controls the spreading distance. Once the spreading has ended (see S7), a new spreading event or avalanche is initiated from another randomly selected single pixel. The critical point separating quiescent and exponentially-growing dynamics, $\sigma_c$, closely follows random graph expectations $\sigma_{rg}(r_0) = N(\sum_{ij} W_{ij})^{-1}$ (see S7)~\cite{newman2018networks}. As $r_0$ goes from large to small values, we observe a transition from MFDP to that of directed percolation on a 2D lattice (Fig.~\ref{fig:modelAnalysis}(a,b))~\cite{munoz1999avalanche}. Fig.~\ref{fig:modelAnalysis}(a,b) also shows that our semi-local analysis can recover the true underlying dynamics. A local analysis agrees only for $r_0 \ll 1$. This is due to the non-locality of the dynamics for larger $r_0$, which allows for propagation farther than neighboring pixels, ``breaking up" avalanches which generates larger effective exponents, explaining the observations in Fig.~\ref{fig:dataAnalysis}(f,g).

More importantly, for $r_0 = 0.85$ we find the correlation length of the exponential fit, $\zeta = 3.7(1)$ (Fig.~\ref{fig:modelAnalysis}(c)), to be consistent with the experimental value $\zeta = 3.5(3)$ (Fig.~\ref{fig:dataAnalysis}(b)). The MLE estimated semi-local and local exponents where found to be: $\tau = 1.43(4)$, $\alpha = 1.73(9)$, $\tau_L =1.91(2)$ and $\alpha_L = 2.8(2)$. The slight differences between model and experiment are likely due to the simplifying assumptions of our model (e.g., homogeneous and isotropic spatial kernel, separation of time scales for avalanches occurrences), which are amplified in a finite system. Other spatial kernels give qualitatively identical results (see S7), such as steeper local exponents and near MFDP semi-local exponents, suggesting a critical spreading process is indeed able to replicate the main features we observe in the experimental data, including the deviations away from true MFDP exponents.

Our analyses of cortical desynchronization on global and localized levels show that incorporating spatial information reveals desynchronization avalanches exhibiting robust scale-free behavior. This re-frames REM/nREM-like dynamics as an absorbing global synchronous state punctuated by localized episodes of desynchronization with no characteristic size or duration. Thus, desynchronization can exist within large clusters of synchronization, but might be obscured without spatial information. The case for localization of REM episodes is further supported by our spreading model. De/synchronization localization may also explain observations of slow-wave sleep during wake periods~\cite{bernardi2015neural}. Overall, our findings connect synchronization and directed percolation, one of the most paradigmatic models of non-equilibrium critical phenomena, for the first time in a biological system. This opens up new avenues of theoretical investigations of sleep(-like) states by way of closely related mathematical theories such as synchronization transitions in spatially extended systems and Reggeon field theories~\cite{PhysRevLett.90.204101, PhysRevE.67.046217}, as well as chimera states, which have been proposed as a mechanism for neural information processing~\cite{panaggio2015chimera, bansal2019cognitive, davidsen2018symmetry, masoliver2022embedded,masoliver2023}.

Due to the similarity of REM dynamics in natural sleep and urethane-anesthesia~\cite{clement2008cyclic}, it is worth considering what role REM localization could play regarding hippocampal-cortical coupling and memory consolidation during natural sleep~\cite{preston2013interplay, rasch2013sleep,buzsaki1989two, colgin2016rhythms, Pedrosa2022.03.08.483425}. A recent study demonstrated increased interaction between the hippocampus and the RSC during REM periods inferred from hippocampal LFP~\cite{de2021hippocampus}. This aligns with the viewpoint that the RSC is a structural hub between the hippocampus~\cite{abadchi2020spatiotemporal} and important cortical areas (primary sensory, prefrontal cortices, etc.). Yet, while we observe high correlations between retrosplenial and hippocampus REM proxies (Fig.~\ref{fig:spatialAnalysis}(d)), desynchronization avalanches were rare here, and more readily occurred in the motor and somatosensory cortex. These regions overlap with areas activated around hippocampal sharp wave ripple complexes~\cite{abadchi2020spatiotemporal}, and negatively correlate with the default mode network of the mouse brain~\cite{whitesell2021regional}, with some deviations in older mice. 
This suggests the mechanism generating large theta-to-delta ratios in the cortex under
urethane anesthesia is different from the one that generates high cortical-hippocampal coupling, motivating future investigations.

Finally, our findings need to be discussed in the context of the critical brain hypothesis (CBH). The CBH posits a self-tuned critical point for brain dynamics that promotes information processing~\cite{kinouchi2006optimal}, with deviations away from criticality reflecting the underlying state of the subject, such as sleep-wake transitions, behavior, or drug induced changes~\cite{scott2014voltage, priesemann2013neuronal, allegrini2015self, bellay2015irregular,meisel2017interplay, yaghoubi2018neuronal, bocaccio2019avalanche, curic2021deconstructing}. This has been motivated at least in part by scale-free statistics in coordinated clusters of causally related neuronal firing events dubbed neuronal avalanches across species and spatial scales~\cite{cocchi2017criticality, PhysRevX.11.021059}. Yet, how critical neuronal avalanches are related to neural oscillations is an open question, and indeed the two initially seem contradictory~\cite{buendia2020hybrid, costa2017self, yang2012maximal}. Theoretical models suggest critical neuronal avalanches require an `edge-of-synchronization' phase transition~\cite{di2018landau}. Our findings provide a complementary view, where criticality is directly associated with (de-)synchronization. Desynchronization spreads in a scale-free manner across the otherwise synchronized cortex, indicative of a critical spreading process and markedly different from neuronal VSD avalanches (see S8). Whether and how these complementary views are connected remains a challenge for the future.

\section*{Acknowledgements}
This work was supported by the Natural Sciences and Engineering Research Council of Canada (DG grants 40352 (MHM) and 05221 (JD)), Alberta Innovates-Technology Futures, and the Killam Trusts (DC).

%apsrev4-2.bst 2019-01-14 (MD) hand-edited version of apsrev4-1.bst
%Control: key (0)
%Control: author (72) initials jnrlst
%Control: editor formatted (1) identically to author
%Control: production of article title (-1) disabled
%Control: page (0) single
%Control: year (1) truncated
%Control: production of eprint (0) enabled
%

%\bibliography{refs}% Produces the bibliography via BibTeX.

\end{document}